# A New Evolutionary Algorithm: Learner Performance Based Behavior Algorithm

Chnoor M. Rahman[1,2], Tarik A. Rashid[3]

[1] Applied Computer Department, College of Medicals and Applied Sciences, Charmo University, Sulaimany, KRG, Iraq.
[2] Technical College of Informatics, Sulaimany Polytechnic University, Sulaimany, KRG, Iraq
[3] Computer Science and Engineering Department, University of Kurdistan Hewler, Erbil, KRG, Iraq.

**ABSTRACT** A novel evolutionary algorithm called learner performance based behavior algorithm (LPB) is proposed in this article. The basic inspiration of LPB originates from the process of accepting graduated learners from high school in different departments at university. In addition, the changes those learners should do in their studying behaviors to improve their study level at university. The most important stages of optimization; exploitation and exploration are outlined by designing the process of accepting graduated learners from high school to university and the procedure of improving the learner's studying behavior at university to improve the level of their study. To show the accuracy of the proposed algorithm, it is evaluated against a number of test functions, such as traditional benchmark functions, CEC-C06 2019 test functions, and a real-world case study problem. The results of the proposed algorithm are then compared to the DA, GA, and PSO. The proposed algorithm produced superior results in most of the cases and comparative in some others. It is proved that the algorithm has a great ability to deal with the large optimization problems comparing to the DA, GA, and PSO. The overall results proved the ability of LPB in improving the initial population and converging towards the global optima. Moreover, the results of the proposed work are proved statistically.

## 1. INTRODUCTION

The computational intelligence (CI) term as a branch of artificial intelligence (AI) was first invented by Bezdek in the early 1990s [1], which motivated a new field to computer-based intelligence. CI in principle consists of any technologies and science-supported approaches for creating, analyzing, and developing intelligent systems [2]. It mainly depends on a set of nature-inspired computational patterns and a numerical collection of data [3]. The study of optimization techniques is one of the main subjects of CI. Optimization is part of any problem that requires decision making, either in economic or engineering fields. Decision-making tasks involve making the best decision to choose between different alternatives. Numerous optimization algorithms exist; however, no single algorithm fits all the different problems. It is crucial for the appropriate optimizer to guarantee that the optimal solution is always reachable. NP-hard problems, for example, are usually not easy to be solved. However, most combinatorial optimization problems, for example, N-Queens, traveling salesperson, and 0/1 Knapsack are NP-hard. To solve this type of problem and depending on the size of the problem, two approaches exist namely; exact methods and metaheuristic methods [4]. Exact methods are useful when the number of decision variables is small. These methods find the optimal solution for the problem. Examples for exact methods are branch and bound algorithm [5], dynamic and linear programming, and so on. The problem with these methods is that they are known as time expensive methods, so that it is not recommended to use them for solving difficult or NP-hard problems. Likewise, where the decision space is discrete or when a large number of decision variables exist, which occurs in most if not in all practical problems of optimization, exact methods cannot show good performance, instead, metaheuristics can be used [4].

Depending on the characteristics, metaheuristic optimization algorithms can be classified in various ways. They can be classified into population-based algorithms and trajectory-based or single-point search algorithms. In the latter case, the algorithm uses a single solution, which means in each iteration only a single solution will manipulate. Hill climbing, tabu search, and simulated annealing are examples of this class of algorithms. On the other hand, population-based algorithms use a population of agents and the whole population is modified in each iteration. Examples for population-based algorithms are genetic algorithm, particle swarm optimization, ant colony optimization, and so on [4].

### 1.1 RELATED WORKS

During the 1960s and 1970s, the metaheuristic optimization algorithms were bloomed. At the beginning of the 1960s, the genetic algorithms (GA) [6] were developed by John Holland and his collaborators. GA is a search technique; it is based on Darwin's theory of evolution and selection of biological systems. The ability of the GA for optimization makes the researchers use it in optimizing a wide range of problems. Since then, it has been modified and hybridized with other techniques to solve various problems. In [7] GA was combined with an active set technique (AST). The hybrid technique was used for optimizing the unsupervised artificial neural network. The aim of this work was to accurately estimate the temperature profiles of the heat conduction model in the head of humans. The results revealed that the hybrid technique produced better and accurate results comparing to standalone approaches, such as GA and AST. Additionally, in [8], GA combined with an interior point technique to optimize a new approach. The approach was solving the initial value of the equation of a Painlev´e II, and its variants, utilizing the feed-forward artificial network. Moreover, in [9], GA combined with IPT to optimize a feed-forward artificial neural network for solving porous fin equation. Better accuracy achieved comparing to other numerical techniques. Similarly, reference [10] designed a neuro-heuristic schema for non-linear second order Thomas-Fermi system. To optimize the schema GA and sequential quadratic programming was utilized. It was discovered that the examined schema was feasible, precise, and effective. Holland's work encouraged many to adopt and develop identical techniques in their research works. Later, in 1966, Fogel et al developed an evolutionary programming technique [11]. In this work, finite state machines were used to represent the solution, and stochastically one of the machines was mutated. Afterward, in 1983, Kirkpatrick et al developed simulated annealing (SA) [12]. SA mimics the process of annealing that utilized for crystallization, which is a physical process in metals and glasses to harden the material. Furthermore, at the beginning of the 1990s, Marco Dorigo completed his Ph.D. thesis on optimization and nature-inspired algorithms. In his thesis, he examined a novel idea known as an ant colony optimization algorithm (ACO) [13]. ACO was inspired by the swarming behavior of social ants utilizing the pheromone to find the source of food and bring the food back to their nest. Later, in 1995, James Kennedy proposed particle swarm optimization (PSO) [14]. PSO can be counted as another significant improvement in the field. It mimics the behaviors of the school of birds or fish. A particle represents a single solution that has a position in the search space. In 2005, Karaboga introduced an artificial bee colony (ABC) [15]. ABC mimics the behaviors of honeybees. It provides well-balanced exploitation and exploration ability. Thereafter, in 2007, Chu and Tsai proposed a new swarm-based optimization algorithm named cat swarm optimization (CSO) algorithm. CSO mimics the behaviors of cats [16]. Yang in 2010 introduced the bat algorithm, which is based on the echolocation behavior of micro bats [17]. In 2014, Mirjalili on the base of hunting behavior, and social hierarchy of grey wolf proposed a new optimization algorithm named as grey wolf optimization (GWO) algorithm [18]. In 2015, the same author proposed the dragonfly algorithm (DA). DA was mainly inspired by the hunting and migrating behaviors of a dragonfly. The latter is called a dynamic (migratory) swarm, and the former is called static (feeding) swarm [19]. Finally, in 2019 fitness dependent optimizer (FDO) developed. It is inspired by the bee swarming reproductive process. FDO mimics the PSO in utilizing velocity to update search agent positions. However, FDO uses the fitness function of the problem to produce weights, and these weights are then used to guide the agents in the exploration and exploitation phase [20].

Since introducing these algorithms for optimization, many researchers utilized them to optimize problems in various fields. However, some other researchers aimed at improving those algorithms. The satisfactory results produced by these algorithms for different optimization problems proved the importance and necessity of them [21-26]. Consequently, researchers continue to propose new algorithms in the field. Many of these algorithms do not have a good balance between exploitation and exploration. Having high exploitation traps the algorithm in the local optimum. Moreover, a high degree of exploration raises the probability of finding global optima but decreases the efficiency. Therefore, having a good balance between exploration and exploitation can make an algorithm perform better compared to the other algorithms [27].

## 1.2 INNOVATIVE CONTRIBUTION

In this paper, a new optimization method, learner performance behavior based algorithm is proposed. The LPB method mimics the process of accepting graduated learners from high school in different colleges and the behaviors of learners that affect their performance during the college study, and the factors that may help the learners to change their high-school study behaviors that are not effective anymore for studying in the college. To implement this, multi-populations can be utilized to demonstrate the learners that have a GPA in different ranges. Consequently, this causes a good balance between exploration and exploitation [27]. The most important features of the proposed work are:

- It is a population based algorithm.
- The initial population is created randomly.
- A percentage of the population is separated.
- The population is divided into a number of sub-populations.
- The highest fitness in the separated group is used to divide the population into sub-populations.
- The sub-populations that contain the best individuals have priority to go through the optimization process first.
- Mutation and crossover operators are used to make changes in the structure of new individuals.

## 1.3 ORGANIZATION

The rest of the paper is organized as follows: Section 2 shows the inspiration of the proposed algorithm. Section 3 presents the features (operators) of GA that are utilized in the proposed technique. The LPB operators and techniques along with the pseudo code are presented in section 4. Furthermore, the results of the algorithm and an inclusive and comparative study on some benchmark test functions along with a real-world problem are presented in section 5. Finally, the conclusion of the work and directions for future researches are shown in section 6.

## 2. INSPIRATION

Every year groups (population) of learners finish their high school and apply to the universities. The applications for some of these learners are accepted and the rest are rejected. Depending on their GPA, the learners are divided into different groups. The process of transferring learners from high school to university starts with a group of graduated learners *M* from high school. Departments from universities specify the number of learners that they want to accept to study in their department. Furthermore, each department specifies the minimum GPA that the learners should have in order to study in that department. This is like grouping the graduated learners from high school *M* to a number of different departments (groups) according to their GPA. The department accepts the applications of the learners if the learner's GPA is in the range of the required GPAs by that department. Among the learners who apply to a specific department, there are a number of learners, which their GPA is under the required GPA. The application for those learners will be rejected. Furthermore, there are learners that their GPA is much higher than the required GPA, thus, these applications have a priority and they will be accepted first, and then the lower ranges, and so on, until the number of accepted applications is equal to the number of learners specified by the department. Furthermore, sometimes it happens that in general, the GPA of learners is low. Thus, some of the departments cannot have a specified number of learners with the required GPA. At these situations before finalizing the list of accepted learners, the department, and the university should decide whether they want to accept learners with less GPA or not.

After accepting the graduated learners from high school in different colleges and departments, the learners will go through a number of difficulties. Because the environment they came from was different from the environment they are in now. In addition, the studying behaviors that they had as high school learners may not be effective anymore. It is normal that many fresh learners are not prepared either academically nor in terms of study skills for college-level study. Working on the learners' studying behaviors, such as seeking help and group working will help them to study more effectively and will result in improving their score during their study in the college [28, 29]. Studying behaviors of a learner can be affected by studying behaviors of learners in the same department or learners in any other department.

The level of learning of transition learners from high school to college can be improved by adopting some effective strategies, which are quite different from those in high school. A number of behaviors have been considered to judge between the strong and weak learners, they include (level of interest, deep processing, effective note taking, problem-solving, group working, seeking help, and self-study). Additionally, according to [30] the learners with a high level of creativity are always strong learners. Depending on the previously mentioned resources, it can be concluded that learners who have a good level of the aforementioned behaviors are good learners.

Moreover, it has been noticed that the quality of metacognition is another key-difference between strong and weak learners. Metacognition refers to the learner's awareness level of understanding of a topic. Those who have poor metacognition are confident and believe that they have done well on exams while they are not and their low score shocks them. When learners get low marks on an exam, they often believe that they should spend more time on studying the subjects. In addition to studying more (although that often helps), however, learners with poor metacognition should change the way they study [31]. Learners with poor metacognition levels usually have poor study strategies, which rise false confidence that they have studied the material well without increasing their actual level of learning. Most fresh learners at colleges have learned to study skills in high school that are no longer effective. They might have a proper sense of metacognition, which accurately informed them when they studied sufficiently during high school, but it is not accurate anymore. This means that entering college requires overcoming the old study strategies with new ones [29, 32]. Besides, having an adequate level of metacognition can cause a good improvement in the learner's level of study and it may have an effect on all the strategies used by the learner. The main inspiration of this algorithm originates from the following steps that a learner goes through:

1) The strategies used to group the learners according to their GPA, and almost all the learners that are accepted in a department have a GPA in a specific range.
2) After accepting the learners in the departments how their studying behaviors can be improved to make them good college learners. The learner's behaviors influenced by each other while they study together.
3) The level of metacognition for learners has a big impact on all the studying behaviors.

In this algorithm, the first step is used to choose individuals from the population. The importance of this step is dividing the main population to some sub-populations and then the individuals will be selected from the sub-populations depending on their fitness. This prevents converge to local optima because selecting individuals will start from the perfect sub-population. The latter two steps are used to improve the individuals by letting the learners work in groups and ask for help from each other. Furthermore, having a good level of metacognition will influence the overall studying behaviours of a learner in a stochastic way (mutation). On the other hand, learners affect the studying behaviours of each other when they study together (crossover).

## 3. GENETIC ALGORITHM OPERATORS

The genetic operators imitate the procedure of the heredity of genes to produce new individuals at each generation. The operators are utilized to make changes in the structure of individuals during the representation. The common genetic operators are crossover, mutation, and selection. Here, we only define crossover and mutation operators.

### 3.1. CROSSOVER

Crossover is the most fundamental genetic operator. It works on two individuals at the same time and produces offspring by integrating features of both individuals. Various crossover techniques are available; however, the most used one is choosing a stochastic cut-point to produce the offspring by integrating the part of one parent to the right of the cut-point with the part of the second parent to the left of the cut-point. For example, one-cut point crossover, two-cut point crossover, multi-cut point crossover, etc. [33].

### 3.2. MUTATION

Mutation creates random changes in different individuals. The simplest form of mutation is altering one or more genes. Mutation in the genetic algorithm has a great role of either a) restoring the lost genes during the selection process, hence, they can be used in another context or b) serving the genes that were not available in the initial population. Various ways of mutation are available for different representations of individuals. For example, uniform mutation, replacement mutation, dynamic mutation, boundary mutation and so on [33].

## 4. LEARNER PERFORMANCE BASED BEHAVIOR ALGORITHM

As the first step in the algorithm, randomly a population of graduated learners *M* who want to apply for different departments in different universities being created. Furthermore, we have an operator and we call it division probability *dp*. As discussed, every department accepts learners that have a GPA greater than or equal to the minimum required GPA. To show this in the algorithm, at first, we use the *dp* parameter to randomly choose a percentage of elements from *M*. Afterwards, we calculate the fitness of each of the chosen individuals and sort them. Then we divide them into two groups, good and bad, depending on their fitness. The former contains the individuals that have a higher GPA and the bad group contains the rest. After this, the fitness of the individuals in the main population *M* is calculated and then filtered. Those individuals that have fitness smaller or equal to the highest fitness in the bad population will be moved to the bad population. The rest of the individuals will be divided into two groups. Those who have fitness smaller or equal to the highest fitness in the good population will be moved to a good population, and those who have fitness higher than the highest fitness in the good population will be moved to the perfect population. Then the number of learners specified by the department will be chosen from the perfect population and good population. If the number of individuals in these two populations was smaller than the number of specified learners by the department this is when the number of learners that have got the required GPA is small and the department should decide whether they do accept other learners with less GPA or not. If they wanted to accept other learners, the rest of the individuals will come from bad population.

After accepting the graduated learners from high school in the departments, as discussed, they may not have effective studying behaviours [29, 32]. However, improving behaviours like help-seeking, group working can have a positive impact on them. In addition, as mentioned in [28, 29] learners can influence each other's behaviour. For example, when they work in groups or when they ask help from each other their studying behaviours will be affected. To show this in the algorithm crossover operator from a genetic algorithm is used. Utilizing a crossover operator will let the individuals exchange some studying behaviours. Consequently, the learner has a set of studying behaviours, which is different from the original studying behaviours owned by the learners. Hence, the overall, behaviours of both individuals will be affected and the produced individuals have different behaviours.

In addition, the level of metacognition has an impact on the overall studying behaviours of a learner. Whenever the metacognition level of a learner is affected, stochastically, the overall studying behaviours of the learner will be affected too [31, 34]. The level of metacognition according to [34] is affected by training the learner using a number of strategies. Using these strategies is excluded from this work. Consequently, the level of metacognition of learners can be affected using a rate that can be specified in the algorithm. As mentioned the metacognition level may affect the overall behaviours of the learners in a stochastic way. So that, randomly changing positions of the behaviours of that individual according to a specific rate or randomly updating the values of studying behaviours of that learner can do this. This is presented in the algorithm by using the mutation operator from the genetic algorithm. Visual 1 shows the pseudo code for LPB.

Definition of symbols:
*M*: the initial random population
*N*: the number of individuals in the new population
*dp*: the percentage of individuals chosen from *M*
*O*: the sub-population chosen from *M* according to the *dp* operator.
*BP:* bad population
*GP*: good population
*PF*: perfect population
*k*: is a counter utilized to count the number of newly created individuals

## 5. RESULTS AND DISCUSSION

In this section, a number of standard benchmark functions in the literature are used to examine LPB. The results are then evaluated against three popular algorithms in the literature: DA, PSO, and GA. The results for 19 classical benchmark functions for PSO, DA, and GA are taken from [19]. Nevertheless, we examined the CEC-C06 2019 test functions to show the ability of the algorithm in solving large scale optimization problems [37]. Additionally, the processing time (PT) in seconds of the algorithm for both groups of the test functions is calculated to show the ability of the algorithm compared to the others in quickly finding the optimal results. Furthermore, to prove the significance of the results, the Wilcoxon rank-sum test [35] is used. Then the algorithm is used to optimize a real-world problem. The parameter settings for LPB are shown in Table 1.

TABLE 1
PARAMETER SETTINGS FOR LPB

| Parameters | Parameter Value |
| --- | --- |
| Crossover rate | 2*round (0.7*population size) |
| Mutation rate | round (0.2*population size) |
| Population Size | 80 |
| dp | 0.5 |

## 5.1. CLASSICAL BENCHMARK TEST FUNCTIONS

To test the performance of the LPB a group of benchmark functions is used. These benchmark functions are divided into three groups: unimodal, multi-modal, and composite test functions [36-39]. Each group has different properties. Unimodal test functions, for example, benchmark the convergence and the exploitation of the algorithm. This group of test functions has a single optimum. However, multi-modal test functions, as their name implies, have multi optimum. They have one global optimum and multi-local optima. To approach the global optimum an algorithm should avoid the entire local optimal solutions. Hence, this group of test functions can benchmark exploration and avoid local optima.

---

**1. [Initialization]**
Randomly create a population $M$
**2. [Specify parameters]**
Specify the number of required learners $N$ for a department, crossover rate and mutation rate
**3. [Create Sub-Populations]**
Use *dp* parameter to randomly choose a percentage of individuals $O$ from $M$
Evaluate the fitness of individuals in $O$
Depending on their fitness, sort the individuals in $O$ (descending order), use one of the sorting methods
Divide $O$ to two populations, good (individuals with high fitness) and bad (individuals with low fitness)
**While** termination condition is not met
    Use *dp* parameter to randomly choose a percentage of individuals $O$ from $M$
    Evaluate the fitness of individuals in $O$
    Depending on their fitness, sort the individuals in $O$ (descending order), use one of the sorting methods
    Divide $O$ to two populations, good (individuals with high fitness) and bad (individuals with low fitness)
    Find fitness for all individuals in the population $M$
    Find the highest fitness in good and bad populations
    **if** an individual from $M$ has fitness <= highest fitness in the bad population
        Move it to the bad population $BP$
    **else if** an individual from $M$ has fitness <= highest fitness in the good population
        Move it to the good population $GP$
    **else**
        Move it to the perfect population $PF$
    **end if**
    **while** $k <= N$
        **if** $PF$ is not empty
            Select an individual from $PF$
        **else if** $GP$ is not empty
            Select an individual from $GP$
        **else**
            Select an individual from $BP$
        **end if**
        $k = k+1$;
    **end while**
    4. Crossover
    5. Mutation
    6. [Termination]
    Repeat the procedure from step 3 until termination condition is met.
**end while**
**7. [Optimal Solution]**
Select the best solution from the perfect population

VISUAL 1: PSEUDO CODE FOR LPB

Finally, the composite test functions are mostly combined, biased, rotated, and shifted versions of the aforementioned groups [39]. They demonstrate the difficulties exist in the real search spaces by providing a huge number of local optima and diverse shapes for various regions. This type of benchmark functions can benchmark the combined exploitation and exploration of an algorithm. See Appendix A, Tables (6-8) for more information about the test functions and their conditions [19]. Ultimately, for each algorithm in Table 2, the test functions are solved 30 times, 80 search agents are utilized over 500 iterations. The average and standard deviation are then calculated. Parameters for GA, PSO, and DA are discussed in reference [19]. For all test functions in Table 1, $dp$ is set to 0.5. The average and standard deviation of the optimal solution is calculated in the last iteration. These two metrics are used to evaluate the overall performance of the algorithms, and to show the stability degree of the algorithms to solve the test functions.

TABLE 2
COMPARISON OF RESULTS OF THE CLASSICAL BENCHMARK FUNCTION BETWEEN LPB, DA, PSO, AND GA

| Test Function | | LPB | DA | PSO | GA |
|---|---|---|---|---|---|
| TF1 | Ave. | 0.001877545 | **2.85E-18** | 4.2E-18 | 748.5972 |
| | Std. | 0.002093616 | **7.16E-18** | 1.31E-18 | 324.9262 |
| | PT (Seconds) | 160.840946 | 1445.243327 | 249.665030 | 65.422226 |
| TF2 | Ave. | 0.005238111 | **1.49E-05** | 0.003154 | 5.971358 |
| | Std. | 0.003652512 | **3.76E-05** | 0.009811 | 1.533102 |
| | PT (Seconds) | 169.076368 | 1259.496468 | 3.826913 | 55.040008 |
| TF3 | Ave. | 36.4748883 | **1.29E-06** | 0.001891 | 1949.003 |
| | Std. | 29.22415523 | **2.1E-06** | 0.003311 | 994.2733 |
| | PT (Seconds) | 202.408611 | 1216.762524 | 12.702411 | 80.126424 |
| TF4 | Ave. | 0.393866 | **0.000988** | 0.001748 | 21.16304 |
| | Std. | 0.135818 | **0.002776** | 0.002515 | 2.605406 |
| | PT (Seconds) | 191.301934 | 1399.014810 | 2.877756 | 63.099468 |
| TF5 | Ave. | 16.76919 | **7.600558** | 63.45331 | 133307.1 |
| | Std. | 22.19251 | **6.786473** | 80.12726 | 85007.62 |
| | PT (Seconds) | 130.846636 | 1707.285731 | 5.224432 | 55.818782 |
| TF6 | Ave. | 0.00203173 | 4.17E-16 | **4.36E-17** | 563.8889 |
| | Std. | 0.0027832 | 1.32E-15 | **1.38E-16** | 229.6997 |
| | PT (Seconds) | 157.547318 | 1550.130722 | 2.795879 | 51.284046 |
| TF7 | Ave. | **0.004975** | 0.010293 | 0.005973 | 0.166872 |
| | Std. | **0.002965** | 0.004691 | 0.003583 | 0.072571 |
| | PT (Seconds) | 158.642028 | 1593.877054 | 8.982616 | 56.555067 |
| TF8 | Ave. | **-3747.65** | -2857.58 | -7.1E+11 | -3407.25 |
| | Std. | **189.0206** | 383.6466 | 1.2E+12 | 164.4776 |
| | PT (Seconds) | 162.354305 | 1738.794894 | 8.266467 | 55.234252 |
| TF9 | Ave. | **0.001567** | 16.01883 | 10.44724 | 25.51886 |
| | Std. | **0.001842** | 9.479113 | 7.879807 | 6.66936 |
| | PT (Seconds) | 159.074029 | 1638.957037 | 4.816792 | 84.833759 |
| TF10 | Ave. | **0.017933** | 0.23103 | 0.280137 | 9.498785 |
| | Std. | **0.013532** | 0.487053 | 0.601817 | 1.271393 |
| | PT (Seconds) | 128.431567 | 1297.325669 | 8.013542 | 84.666823 |
| TF11 | Ave. | **0.066355** | 0.193354 | 0.083463 | 7.719959 |
| | Std. | **0.030973** | 0.073495 | 0.035067 | 3.62607 |
| | PT (Seconds) | 130.664299 | 1210.086084 | 9.429028 | 56.656545 |
| TF12 | Ave. | 2.78659E-05 | 0.031101 | **8.57E-11** | 1858.502 |
| | Std. | 3.83626E-05 | 0.098349 | **2.71E-10** | 5820.215 |
| | PT (Seconds) | 140.837076 | 1464.060419 | 22.898798 | 102.745164 |
| TF13 | Ave. | **0.000309** | 0.002197 | 0.002197 | 68047.23 |
| | Std. | **0.000512** | 0.004633 | 0.004633 | 87736.76 |
| | PT (Seconds) | 139.449467 | 1339.438272 | 16.752814 | 103.377836 |
| TF14 | Ave. | **0.998004** | 103.742 | 150 | 130.0991 |
| | Std. | **1.26E-11** | 91.24364 | 135.4006 | 21.32037 |
| | PT (Seconds) | 170.207352 | 1034.450489 | 86.298548 | 152.142368 |
| TF15 | Ave. | **0.002358** | 193.0171 | 188.1951 | 116.0554 |
| | Std. | **0.003757** | 80.6332 | 157.2834 | 19.19351 |
| | PT (Seconds) | 247.224271 | 1659.652400 | 8.250347 | 54.974533 |
| TF16 | Ave. | **-1.03163** | 458.2962 | 263.0948 | 383.9184 |
| | Std. | **2.46E-06** | 165.3724 | 187.1352 | 36.60532 |
| | PT (Seconds) | 181.858429 | 969.827007 | 4.247415 | 80.998874 |
| TF17 | Ave. | **0.397888** | 596.6629 | 466.5429 | 503.0485 |
| | Std. | **3.16E-06** | 171.0631 | 180.9493 | 35.79406 |
| | PT (Seconds) | 141.213291 | 1018.757437 | 2.607163 | 50.990811 |
| TF18 | Ave. | **3.000142** | 229.9515 | 136.1759 | 118.438 |
| | Std. | **0.000283** | 184.6095 | 160.0187 | 51.00183 |
| | PT (Seconds) | 180.663489 | 1001.716543 | 2.718852 | 80.273981 |
| TF19 | Ave. | **-3.86278** | 679.588 | 741.6341 | 544.1018 |
| | Std. | **9.61E-07** | 199.4014 | 206.7296 | 13.30161 |
| | PT (Seconds) | 169.415055 | 1312.805448 | 8.952319 | 77.905123 |

For each test function in Table 2, superior results are shown in bold. As shown in Table 2, for the first six unimodal test functions (TF1-TF6), the DA algorithm outperforms the LPB, and also PSO performs better in the (TF1-TF6). This proves that the exploitation and the convergence speed of the algorithm are not better than the algorithms used in the comparison. However, the results of the unimodal test functions of the LPB comparing to the GA are evident that LPB has a greater exploitation rate and convergence speed. In addition, LPB outperforms both PSO and DA in the last unimodal test function (TF7) and PSO in TF5 as well. Nevertheless, the LPB provides better results than the other algorithms in all the other test functions. PSO, however, provided a better result in TF12. These results show the ability of the branching algorithm in avoiding local optima, exploring the search space, and balancing exploration and exploitation. Results of the test functions TF7-TF19 proved that LPB has a superior exploration and a perfect ability in avoiding local optima, and also it has a superior balance between exploration and exploitation phases comparing to the DA, PSO, and GA. As shown in Table 2, it can be concluded that the LPB has the first rank among the other algorithms because it outperforms the other algorithms in 12 functions out of 19 functions. Fig. 1 shows the convergence curve for the proposed algorithm. In Fig. 1, for each group of the test functions, one function is selected (F2 for unimodal, F9 for multi-modal, and F17 for composite test functions), and cost refers to the fitness value for the global best.

For the traditional benchmark functions, the PT of the LPB is much smaller comparing to the DA. The reason for this is that in the first stage of the LPB, a subset of the population is chosen based on this smaller group other subpopulations are built. The perfect subpopulation has priority to be optimized first, then the good subpopulation and so on. Since the subpopulations are much smaller compared to the main population, searching for the solutions in these subpopulations is speeder. This improves the randomness and saves the optimization time simultaneously. However, compared to the PSO and GA, the PT of the LPB is higher.

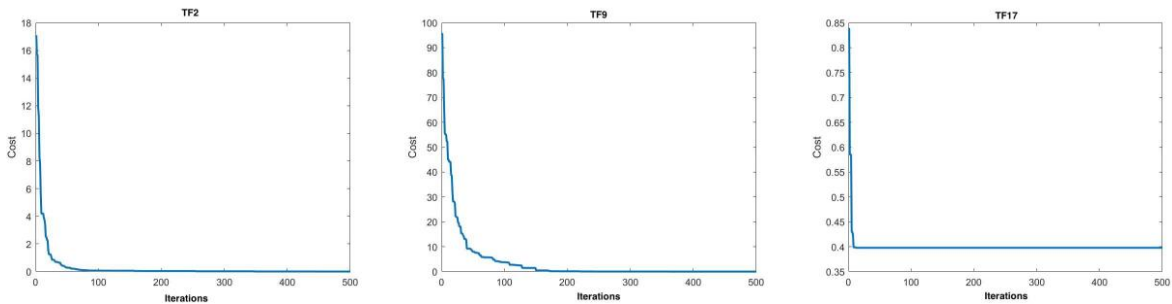

FIGURE 1: Convergence curve for LPB on unimodal, multi-modal, and composite benchmark function

*5.2. CEC-C06 2019 BENCHMARK TEST FUNCTIONS*

Many real-world problems exist in which time is not as important as getting an accurate answer. In addition, practically people tune an algorithm and execute it more than one trail if they wanted. This means users try to find the most successful algorithm for their scenario regardless of time. It is this feature of numerical optimization, which the CEC-C06 benchmark test functions also known as "The 100-digit challenge" examine. They calculate the values of functions at "horizontal" slices of the convergence plot [39]. These test functions are considered for use in an annual competition of optimization. They are used to evaluate the algorithm for large scale optimization problems. The first three functions, CEC01 to CEC03, have various dimensions as shown in Appendix B Table 9. On the other hand, the CEC04 to CEC10 functions set as 10-dimensional minimization problems in the range [-100, 100], and they are shifted and rotated. All the CEC functions are scalable and all global optimum of these functions were united towards point 1. The results of the CEC-C06 2019 test functions for the LPB, DA, and PSO are shown in Table 3. For each test function in Table 3, superior results are shown in bold. The test functions are solved 30 times utilizing 80 search agents over 500 iterations. The average, standard deviation, and processing time are then calculated. The results of the CEC-C06 2019 benchmark functions for DA and PSO are taken from [40]. As shown in Table 3, the value of metrics, average, and standard deviation for the LPB algorithm in almost all the CEC-C06 2019 test functions are smaller than DA, and PSO. However, PSO showed its superiority in CEC04. Additionally, the results of the LPB and PSO for optimizing CEC05, and CEC09 are comparative. The results of the CEC-C06 2019 benchmark functions revealed that for large scale optimization problems LPB provides better results compared to the DA, and PSO.

The processing time for the LPB and DA for the CEC-C06 2019 is also shown in Table 3. As clear, the PT for the LPB for optimizing all the functions is much smaller. The reason for this, as mentioned earlier, is that in the first stage of the LPB, a subset of the population is chosen based on this smaller group other subpopulations are built. The perfect subpopulation has priority to be optimized first, then the good subpopulation and so on. Since the subpopulations are much smaller compared to the main population, searching for the solutions in these subpopulations is speeder. Consequently, this improves the randomness and saves the optimization time simultaneously. However, compared to the PSO and GA, the PT of the LPB is higher.

TABLE 3
IEEE CEC 2019 BENCHMARK TEST RESULTS.

| CEC Function | | LPB | DA | PSO |
|---|---|---|---|---|
| CEC01 | Ave. | **7494381363.65768** | $543 \times 10^8$ | 1.47127E+12 |
| | Std. | **8138223463.28023** | $669 \times 10^8$ | 1.32362E+12 |
| | PT (Seconds) | 377.373846 | 2034.958870 | 382.330436 |
| CEC02 | Ave. | **17.63898** | 78.0368 | 15183.91348 |
| | Std. | **0.31898** | 87.7888 | 3729.553229 |
| | PT (Seconds) | 140.912536 | 2122.108475 | 6.064791 |
| CEC03 | Ave. | **12.7024** | 13.7026 | 12.70240422 |
| | Std. | **0** | 0.0007 | 9.03E-15 |
| | PT (Seconds) | 144.194876 | 2223.799974 | 8.901970 |
| CEC04 | Ave. | 77.90824 | 344.3561 | **16.80077558** |
| | Std. | 29.88519 | 414.0982 | **8.199076134** |
| | PT (Seconds) | 137.305797 | 1720.974833 | 5.179151 |
| CEC05 | Ave. | **1.18822** | 2.5572 | **1.138264955** |
| | Std. | **0.10945** | 0.3245 | **0.089389848** |
| | PT (Seconds) | 138.406681 | 1722.243949 | 5.370252 |
| CEC06 | Ave. | **3.73895** | 9.8955 | 9.305312443 |
| | Std. | **0.82305** | 1.6404 | 1.69E+00 |
| | PT (Seconds) | 142.041586 | 1401.682147 | 131.167162 |
| CEC07 | Ave. | **145.28775** | 578.9531 | 160.6863065 |
| | Std. | **177.8949** | 329.3983 | 104.2035197 |
| | PT (Seconds) | 122.135692 | 1376.289834 | 5.436392 |
| CEC08 | Ave. | **4.88769** | 6.8734 | 5.224137165 |
| | Std. | **0.67942** | 0.5015 | 0.786760649 |
| | PT (Seconds) | 138.207450 | 1802.883649 | 5.527832 |
| CEC09 | Ave. | **2.89429** | 6.0467 | **2.373279266** |
| | Std. | **0.23138** | 2.871 | **0.018437068** |
| | PT (Seconds) | 141.699472 | 1365.799778 | 4.446880 |
| CEC10 | Ave. | **20.00179** | 21.2604 | 20.28063455 |
| | Std. | **0.00233** | 0.1715 | 0.128530895 |
| | PT (Seconds) | 147.995515 | 1699.088096 | 9.462923 |

### 5.3. STATISTICAL TESTS

The Wilcoxon rank-sum test function [35] is used to verify the importance of the results statistically. The *p* values reported in Table 4 for classical benchmark test functions prove that for most of the test functions the LPB showed significantly better results compared to the DA. Again, in reference [19] it was proved that the results of the DA are statistically significant comparing to PSO and GA. This means there is no need to compare the proposed algorithm with PSO and GA statistically since it proves its superiority against DA. As shown in Table 4, all the results except (TF6, TF11, TF12, and TF19) were smaller than 0.05, which proves the importance of the results of the proposed algorithm.

### 5.4. REAL WORLD APPLICATION

In this section, the algorithm is used to optimize a generalized assignment problem. The problem and its representation are discussed in the following two sections.

### 5.4.1. PROBLEM DEFINITION

A generalized assignment problem known as (GAP) is a popular NP-hard combinatorial optimization problem [41]. The main goal in the GAP is assigning a set of tasks to a set of workers with minimum cost. In this work, we assign cases in the court to justice teams in a way that the cases could be finished within a minimum number of working hours. Assigning cases and justice administration in the judicial system is routine works, however, they are very time-consuming. Increasing caseloads at any time will make the problem more series. In this work, we use the proposed algorithm to assign the right case to the right justice team and to assign a proper time to deliver the decision of the court. The cases should be assigned to the teams in the base of the number of hours required by that team to deal with that case. So that, it can be considered that *N* cases and *N* justice teams are available where we have to assign each case to one and only one justice team in a way that the total hours of assigning cases to the justice teams are minimized. To form the problem mathematically, first we define the following symbols:

i → row number indicating $i^{th}$ case  $\qquad$ i ε [1, N]
j → column number indicating $j^{th}$ justice team  $\qquad$ j ε [1, N]

$C[i][j]$ → cost of allocating $i^{th}$ case to the $j^{th}$ team
$X[i][j] = 1$ if $j^{th}$ justice team is assigned to $i^{th}$ case
$X[i][j] = 0$ otherwise.

The problem can be formulated mathematically as:

$$Min \sum_{i=1}^{N} \sum_{j=1}^{N} C[i][j]X[i][j] \qquad (3)$$

Subject to:

$$\sum_{i=1}^{N} X[i][j] = 1, \forall\, i \in N = \{1,2,\ldots N\}$$

$$\sum_{j=1}^{N} X[i][j] = 1, \forall\, j \in N = \{1,2,\ldots N\}$$

$X[i][j] \in \{0,1\}$

TABLE 4
THE WILCOXON RANK-SUM TEST OVERALL RUNS FOR CLASSICAL BENCHMARK TEST FUNCTIONS

| Test Function | LPB Vs. DA |
|---|---|
| TF1 | **7.72E-06** |
| TF2 | **1.07E-10** |
| TF3 | **5.52E-09** |
| TF4 | **3.42E-06** |
| TF5 | **0.006739** |
| TF6 | 0.75328294 |
| TF7 | **7.77E-13** |
| TF8 | **4.23E-27** |
| TF9 | **1.91E-05** |
| TF10 | **1.08E-09** |
| TF11 | **5.96E-17** |
| TF12 | 0.138213 |
| TF13 | 0.185156 |
| TF14 | **0.04631** |
| TF15 | **0.025386** |
| TF16 | **0.033765** |
| TF17 | **0.089253** |
| TF18 | **0.007899** |
| TF19 | 0.35758 |

### 5.4.2. PROBLEM REPRESENTATION

Representing the problem will be a row from 1 to *N* examining the square cost matrix. Every individual in the population is a permutation from 1 to *N*. If the $j^{th}$ element in the row is $i$, thus, the $i^{th}$ case will be given to the $j^{th}$ justice team. For instance, let's consider the following matrix:

|  | Team1 | Team2 | Team3 | Team4 | Team5 |
|---|---|---|---|---|---|
| Case1 | 23 | 21 | 12 | 30 | 19 |
| Case2 | 30 | 25 | 13 | 22 | 21 |
| Case3 | 21 | 23 | 32 | 40 | 15 |
| Case4 | 12 | 32 | 40 | 32 | 29 |
| Case5 | 20 | 15 | 21 | 27 | 22 |

If the solution is [4 5 2 3 1] that means case 4 in column 1 with cost 12 will be given to the first justice team, case 5 in column 2 with cost 23 will be given to the second team, case 2 with cost 13 in column 3 will be given to the third team and so on. Because of the constraint that says every case should be assigned to one and only one team and according to the encoding used, elements in each tuple should be unique. Thus, partially mapped crossover [42] was used where the individuals are permutations of numbers between 1 and *N*. For mutation, swap mutation was used, randomly two points between 1 and *N* were generated and values of those two positions were swapped. The proposed algorithm was applied to the problem using 80 individuals, for 200 iterations. To verify the ability of the algorithm to solve the problem different size of the matrix was given to the algorithm, as shown in Table 5. To run the program a standard laptop with processor Intel Core i7, 16 GHz was used. The results for different matrix sizes are shown in Table 5.

TABLE 5
RESULT OF THE COURT CASE ASSIGNMENT PROBLEM WITH VARYING SIZE.

| Size of matrix | Optimal Solution | No. Of Generations | Time Required (Sec.) |
|---|---|---|---|
| 10x10 | 218 | 17 | 0.14 |
| 15x15 | 350 | 15 | 0.17 |
| 20x20 | 425 | 34 | 0.33 |
| 30x30 | 676 | 57 | 0.53 |

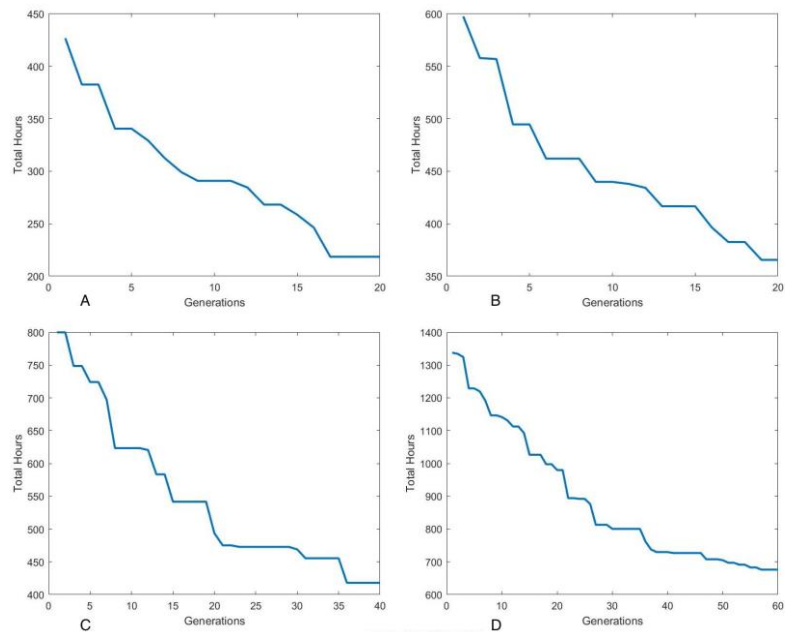

FIGURE 2: Convergence to the global minimum using a different number of cases and justice teams, A) 10x10, B) 15x15, C) 20x20, D) 30x30

In all the cases the population size was kept to 80, and the values for the matrix were generated in the range [10 100]. Fig. 2 shows the convergence of the algorithm towards the global minimum for solving the aforementioned problem using a different size for the matrix.

## 6. CONCLUSIONS

This paper proposed another metaheuristic algorithm based on the process of transferring graduated learners from high school to university and improving the studying behaviors of the learners at colleges. The genetic algorithm inspired this algorithm. The two most important phases of metaheuristic algorithms (exploitation and exploration) were outlined. Mimicking the process of transferring graduated learners from high school to college and dividing them into different groups according to their GPA outlined the former phase. The exploration phase, however, was designed by mimicking the process of improving the level of learners by utilizing various affective study skills. The parameters used in the LPB were *dp*, crossover, mutation. The *dp* parameter is used in the first steps of the algorithm to divide the population into different groups. The latter two parameters were utilized in the process of improving learners studying skills.

The ability of the proposed work was benchmarked using traditional test function and the CEC-C06 2019 functions. The results were compared to PSO, GA and one of the most recently developed algorithms, which is DA. It was proven that the LPB performed better in most of the cases. Moreover, The processing time of the algorithm was compared to the GA, PSO, and DA. the PT of the proposed work was much smaller compared to the DA. However, it was found that the processing time of the PSO, and GA are smaller than the LPB. Additionally, the Wilcoxon rank-sum test function was used to prove the significance of the results. Furthermore, the ability of the algorithm was tested using a real-world NP-hard problem. Again, the results proved the effectiveness of the proposed algorithm in solving a real-world problem. As per finding of the examined work, it can be concluded that the proposed work is able to outperform most of the algorithms in the literature. However, bigger problem sizes for combinatorial optimization could be a challenge for LPB. Therefore, it is recommended for researchers in different fields to use it as an optimization technique.

For future works, a number of research directions can be recommended. First of all, the authors will focus on reducing the processing time of the algorithm. Moreover, implementing the multi objective version of the algorithm is another research direction. Modifying the algorithm to improve the exploitation phase of LPB is another area that the authors are planning to implement in the future. Moreover, another future work is finding new parameters to replace the parameters from the genetic algorithm. In addition, utilizing the proposed technique to optimize different problems and compare the results with other heuristic techniques.

## Acknowledgment

The authors would like to send special thanks to Mr. Ahmed Saadaldin Qosaeri from the University of Kurdistan Hewler (UKH), for his thoughtful ideas and discussion.

## APPENDICE

### APPENDIX A

Single-objective test problems are used in this work. See Tables 6, 7, and 8 for the mathematical representation of traditional benchmark functions used in this work.

TABLE 6
UNIMODAL BENCHMARK FUNCTIONS

| Function | Dimension | Range | Shift position | $f_{min}$ |
|---|---|---|---|---|
| $TF1(x) = \sum_{i=1}^{n} x_i^2$ | 10 | [-100, 100] | [-30, -30, … -30] | 0 |
| $TF2(x) = \sum_{i=1}^{n} |x_i| + \prod_{i=1}^{n} |x_i|$ | 10 | [-10,10] | [-3, -3, … -3] | 0 |
| $TF3(x) = \sum_{i=1}^{n} \left( \sum_{j=1}^{i} x_j \right)^2$ | 10 | [-100, 100] | [-30, -30, … -30] | 0 |
| $TF4(x) = \max_i \{|x|, 1 \leq i \leq n\}$ | 10 | [-100, 100] | [-30, -30, … -30] | 0 |
| $TF5(x) = \sum_{i=1}^{n-1} [100 (x_{i+1} - x_1^2)^2 + (x_i - 1)^2]$ | 10 | [-30, 30] | [-15, -15, … -15] | 0 |
| $TF6(x) = \sum_{i=1}^{n} ([x_i + 0.5])^2$ | 10 | [-100, 100] | [-750, … -750] | 0 |
| $TF7(x) = \sum_{i=1}^{n} ix_i^4 + random[0,1]$ | 10 | [-1.28, 1.28] | [-0.25, … -0.25] | 0 |

TABLE 7
MULTI-MODAL BENCHMARK FUNCTIONS

| Function | Range | Shift Position | $f_{min}$ |
|---|---|---|---|
| $TF8(x) = \sum_{i=1}^{n} -x_i^2 \sin\left(\sqrt{|x_i|}\right)$ | [-500, 500] | [-300, … -300] | -418.9829 X 5 |
| $TF9(x) = \sum_{i=1}^{n} [x_i^2 - 10 \cos(2\pi x_i) + 10]$ | [-5.12, 5.12] | [-2, -2, … -2] | 0 |
| $TF10(x) = -20exp\left(-0.2\sqrt{\sum_{i=1}^{n} x_i^2}\right) - exp\left(\frac{1}{n}\sum_{i=1}^{n} \cos(2\pi x_i)\right) + 20 + e$ | [-32, 32] |  | 0 |
| $TF11(x) = \frac{1}{4000}\sum_{i=1}^{n} x_i^2 - \prod_{i=1}^{n} \cos\left(\frac{x_i}{\sqrt{i}}\right) + 1$ | [-600, 600] | [-400, … -400] | 0 |

$$TF12(x) = \frac{\pi}{n}\left\{10\sin(\pi y_1) + \sum_{i=1}^{n-1}(y_i - 1)^2[1 + 10\sin^2(\pi y_i + 1)]\right.$$
$$\left. + (y_n - 1)^2\right\} + \sum_{i=1}^{n} u(x_i, 10, 100, 4)$$
[-50, 50]   [-30, 30, ... 30]   0

$$TF13(x) = 0.1\left\{\sin^2(3\pi x1)\right.$$
$$+ \sum_{i=1}^{n}(x_i - 1)^2[1 + \sin^2(3\pi x_i + 1)]$$
$$\left. + (x_n - 1)^2[1 + \sin^2(2\pi x_n)]\right\}$$
$$+ \sum_{i=1}^{n} u(x_i, 5, 100, 4)$$
[-50, 50]   [-100, ... -100]   0

TABLE 8
COMPOSITE BENCHMARK FUNCTIONS

| Function | Dimension | Range | $f_{min}$ |
|---|---|---|---|
| $TF14(CF1)$<br>$f1, f2, f3 \dots f10 = Sphere\ function$<br>$\delta1, \delta2, \delta3 \dots \delta10 = [1,1,1, \dots 1]$<br>$\lambda1, \lambda2, \lambda3, \dots \lambda10 = [\frac{5}{100}, \frac{5}{100}, \frac{5}{100}, \dots \frac{5}{100}]$ | 10 | [-5, 5] | 0 |
| $TF15(CF2)$<br>$f1, f2, f3 \dots f10 = Grienwank's\ function$<br>$\delta1, \delta2, \delta3 \dots \delta10 = [1,1,1, \dots 1]$<br>$\lambda1, \lambda2, \lambda3, \dots \lambda10 = [\frac{5}{100}, \frac{5}{100}, \frac{5}{100}, \dots \frac{5}{100}]$ | 10 | [-5, 5] | 0 |
| $TF16(CF3)$<br>$f1, f2, f3 \dots f10 = Grienwank's\ function$<br>$\delta1, \delta2, \delta3 \dots \delta10 = [1,1,1, \dots 1]$<br>$\lambda1, \lambda2, \lambda3, \dots \lambda10 = [1, 1, 1, \dots 1]$ | 10 | [-5, 5] | 0 |
| $TF17(CF4)$<br>$f1, f2 = Ackley's function$<br>$f3, f4 = Rastrigin's function$<br>$f5, f6 = Weierstrass's function$<br>$f7, f8 = Griewank's function$<br>$f9, f10 = Sphere\ function$<br>$\delta1, \delta2, \delta3 \dots \delta10 = [1,1,1, \dots 1]$<br>$\lambda1, \lambda2, \lambda3, \dots \lambda10 = [\frac{5}{32}, \frac{5}{32}, 1, 1, \frac{5}{0.5}, \frac{5}{0.5}, \frac{5}{100}, \frac{5}{100}, \frac{5}{100}, \frac{5}{100}]$ | 10 | [-5, 5] | 0 |
| $TF18(CF5)$<br>$f1\ f2 = Rastrigin's function$<br>$f3, f4 = Weierstrass's function$<br>$f5, f6 = Griewank's function$<br>$f7, f8 = Ackley's function$<br>$f9, f10 = Sphere\ function$<br>$\delta1, \delta2, \delta3 \dots \delta10 = [1,1,1, \dots 1]$<br>$\lambda1, \lambda2, \lambda3, \dots \lambda10 = [\frac{1}{5}, \frac{1}{5}, \frac{5}{0.5}, \frac{5}{0.5}, \frac{5}{100}, \frac{5}{100}, \frac{5}{32}, \frac{5}{32}, \frac{5}{100}, \frac{5}{100}]$ | 10 | [-5, 5] | 0 |
| $TF19(CF6)$<br>$f1\ f2 = Rastrigin's function$<br>$f3, f4 = Weierstrass's function$<br>$f5, f6 = Griewank's function$<br>$f7, f8 = Ackley's function$<br>$f9, f10 = Sphere\ function$<br>$\delta1, \delta2, \delta3 \dots \delta10 = [0.1, 0.2, 0.3, 0.4, 0.5, 0.6, 0.7, 0.8, 0.9, 1]$ | 10 | [-5, 5] | 0 |

$$\lambda 1, \lambda 2, \lambda 3, \dots \lambda 10 = [0.1 * \frac{1}{5}, 0.2 * \frac{1}{5}, 0.3 * \frac{5}{0.5}, 0.4 * \frac{5}{0.5}, 0.5 * \frac{5}{100}, 0.6 * \frac{5}{100}, 0.7 * \frac{5}{32}, 0.8 * \frac{5}{32}, 0.9 * \frac{5}{100}, 1 * \frac{5}{100}]$$

## APPENDIX B

The CEC-C06 2019 benchmark functions are shown in the following table:

TABLE 9
CEC-C06 2019 BENCHMARK FUNCTIONS [37]

| Function | Functions | Dimension | Range | $f_{min}$ |
|---|---|---|---|---|
| CEC01 | STORN'S CHEBYSHEV POLYNOMIAL FITTING PROBLEM | 9 | [-8192, 8192] | 1 |
| CEC02 | INVERSE HILBERT MATRIX PROBLEM | 16 | [-16384, 16384] | 1 |
| CEC03 | LENNARD-JONES MINIMUM ENERGY CLUSTER | 18 | [-4, 4] | 1 |
| CEC04 | RASTRIGIN'S FUNCTION | 10 | [-100, 100] | 1 |
| CEC05 | GRIENWANK'S FUNCTION | 10 | [-100, 100] | 1 |
| CEC06 | WEIERSRASS FUNCTION | 10 | [-100, 100] | 1 |
| CEC07 | MODIFIED SCHWEFEL'S FUNCTION | 10 | [-100, 100] | 1 |
| CEC08 | EXPANDED SCHAFFER'S F6 FUNCTION | 10 | [-100, 100] | 1 |
| CEC09 | HAPPY CAT FUNCTION | 10 | [-100, 100] | 1 |
| CEC10 | ACKLEY FUNCTION | 10 | [-100, 100] | 1 |